\documentclass[%
 aip,
 amsmath,amssymb,
preprint,
]{revtex4-1}
\usepackage{graphicx}
\usepackage{dcolumn}
\usepackage{bm}
\usepackage[utf8]{inputenc}
\usepackage[T1]{fontenc}
\usepackage{mathptmx}
\usepackage{etoolbox}
\usepackage{soul}
\usepackage{gensymb}
\usepackage[caption=false]{subfig}
\usepackage{multirow}
\usepackage{threeparttable}
\usepackage{color}
\usepackage[dvipsnames]{xcolor}
\usepackage{chemformula}
\usepackage{xspace}
\usepackage{orcidlink}
\makeatletter
\def\@email#1#2{%
 \endgroup
 \patchcmd{\titleblock@produce}
  {\frontmatter@RRAPformat}
  {\frontmatter@RRAPformat{\produce@RRAP{*#1\href{mailto:#2}{#2}}}\frontmatter@RRAPformat}
  {}{}
}%
\makeatother
\def\Hext{\textmu\textsubscript{0}H\textsubscript{ext }}
\def\Ex{E$\textsubscript{x } $}
\def\Ey{E$\textsubscript{y } $}
\def\EX{E$\textsubscript{x}$}

\def\wone{$\omega_{1 } $ }
\def\wtwo{$\omega_{2 } $ }
\def\wthree{$\omega_{3 } $ }
\def\Neel{Ne\'el }
\usepackage{amsmath} 

\begin{document}
\preprint{AIP/123-QED}
\title[Separating terahertz spin and charge contributions from ultrathin antiferromagnetic heterostructures]{Separating terahertz spin and charge contributions from ultrathin antiferromagnetic heterostructures}
\author{\orcidlink{0000-0001-7052-3016}T.~W.~J.~Metzger}
    \email{thomas.metzger@ru.nl}
     \affiliation{Institute for Molecules and Materials, Radboud University, 6525 AJ Nijmegen, The Netherlands}

\author{
P.~Fischer}
    \affiliation{Department of Physics and Center for Applied Photonics, University of Konstanz, D-78457 Konstanz, Germany
    }

\author{\orcidlink{0000-0002-7789-604X}T.~Kikkawa}
\affiliation{Department of Applied Physics, The University of Tokyo, Tokyo 113-8656, Japan}

\author{\orcidlink{0000-0003-2675-0718} E.~Saitoh}
\affiliation{Department of Applied Physics, The University of Tokyo, Tokyo 113-8656, Japan}
\affiliation{Institute for AI and Beyond, The University of Tokyo, Tokyo 113-8656, Japan}
\affiliation{WPI Advanced Institute for Materials Research, Tohoku University, Sendai 980-8577, Japan}
\affiliation{RIKEN Center for Emergent Matter Science (CEMS), Wako 351–0198, Japan}

 \author{\orcidlink{0000-0002-0709-042X}A.~V.~Kimel}
     \affiliation{Institute for Molecules and Materials, Radboud University, 6525 AJ Nijmegen, The Netherlands}

\author{\orcidlink{0000-0003-3537-497X}D.~Bossini}
    \affiliation{Department of Physics and Center for Applied Photonics, University of Konstanz, D-78457 Konstanz, Germany
    }
\date{\today}
\begin{abstract}
Femtosecond laser excitation of nanometer thin heterostructures comprising a heavy metal and a magnetically ordered material is known to result in the emission of terahertz radiation. However, the nature of the emitted radiation from heavy metal~/~antiferromagnet heterostructures has sparked debates and controversies in the literature. Here, we unambiguously separate spin and charge contributions from Pt~/~NiO heterostructures by introducing an unprecedented methodology combining high external magnetic fields with a symmetry analysis of the emitted terahertz polarization. We observe two distinct mechanisms of terahertz emission which we identify as optical difference frequency generation and ultrafast laser-induced quenching of the magnetization. We emphasize the absence of spin transport effects and signatures of coherent magnons. Overall, our work provides a general experimental methodology to separate spin and charge contributions to the laser-induced terahertz emission from heterostructures comprising a magnetically ordered material thus holding great potential for advancing terahertz spintronics and establishing terahertz orbitronics. 
\end{abstract}
\keywords{Antiferromagnets, ultrafast demagnetization, NIR pump - terahertz emission probe spectroscopy}
\maketitle
Antiferromagnetic (AF) spintronics is highly promising for fundamental research and future applications. In particular, the potential of antiferromagnets lies in their robustness against perturbing external electromagnetic fields and in their dynamics, intrinsically faster than in ferromagnets, as the magnetic eigenfrequencies enter the THz range~\cite{nemec_antiferromagnetic_2018,jungwirth_antiferromagnetic_2016,baltz_antiferromagnetic_2018,jungfleisch_perspectives_2018,walowski_perspective_2016}. The key to antiferromagnetic spintronics is the ability to convert the spin dynamics into a charge signal. For this purpose, heterostructures comprising a heavy metal and an antiferromagnet (HM~/~AF) have been studied recently~\cite{kholid_importance_2022,schmoranzerova_thermally_2023,qiu_terahertz_2023}. In particular, thin film Pt/NiO heterostructures have become a model system in antiferromagnetic spintronics. Laser-induced THz emission as a result of spin-current injection from NiO~\cite{qiu_ultrafast_2021}, optically triggered torque on NiO spins as a consequence of femtosecond laser heating of Pt~\cite{rongione_emission_2023} and magneto-optical pump-probe experiments investigating the ultrafast demagnetization of antiferromagnetic NiO sublattices~\cite{wust_indirect_2022} were reported.
%
%
This rich variety of the observed and sometimes discrepant effects in the literature sparks controversy and arises from the complexity of these heterostructures. In principle, both layers (NiO and Pt), and even the interface between them can be sources of signal. In this intrinsically convoluted scenario, assessing the microscopic origin of the laser-induced THz emission is challenging. Separating magnetic from non-magnetic contributions requires  measurements demonstrating a characteristic change upon crossing the ordering temperature or external magnetic field, intense enough to modify the spin configuration in the ground state. Such measurements have hitherto not been reported. In fact, the studies in the literature were conducted either without an externally applied magnetic field or in a field, which was too weak to modify the AF spin ordering. Moreover, the temperature dependence of the THz emission was investigated only in the antiferromagnetic phase, far away from the critical temperature. In this scenario, we argue that a conclusive and robust proof of the magnetic nature of laser-induced THz emission signals at the picosecond timescale, is still missing.
%

\begin{figure}
    \centering
    \includegraphics[width=\columnwidth]{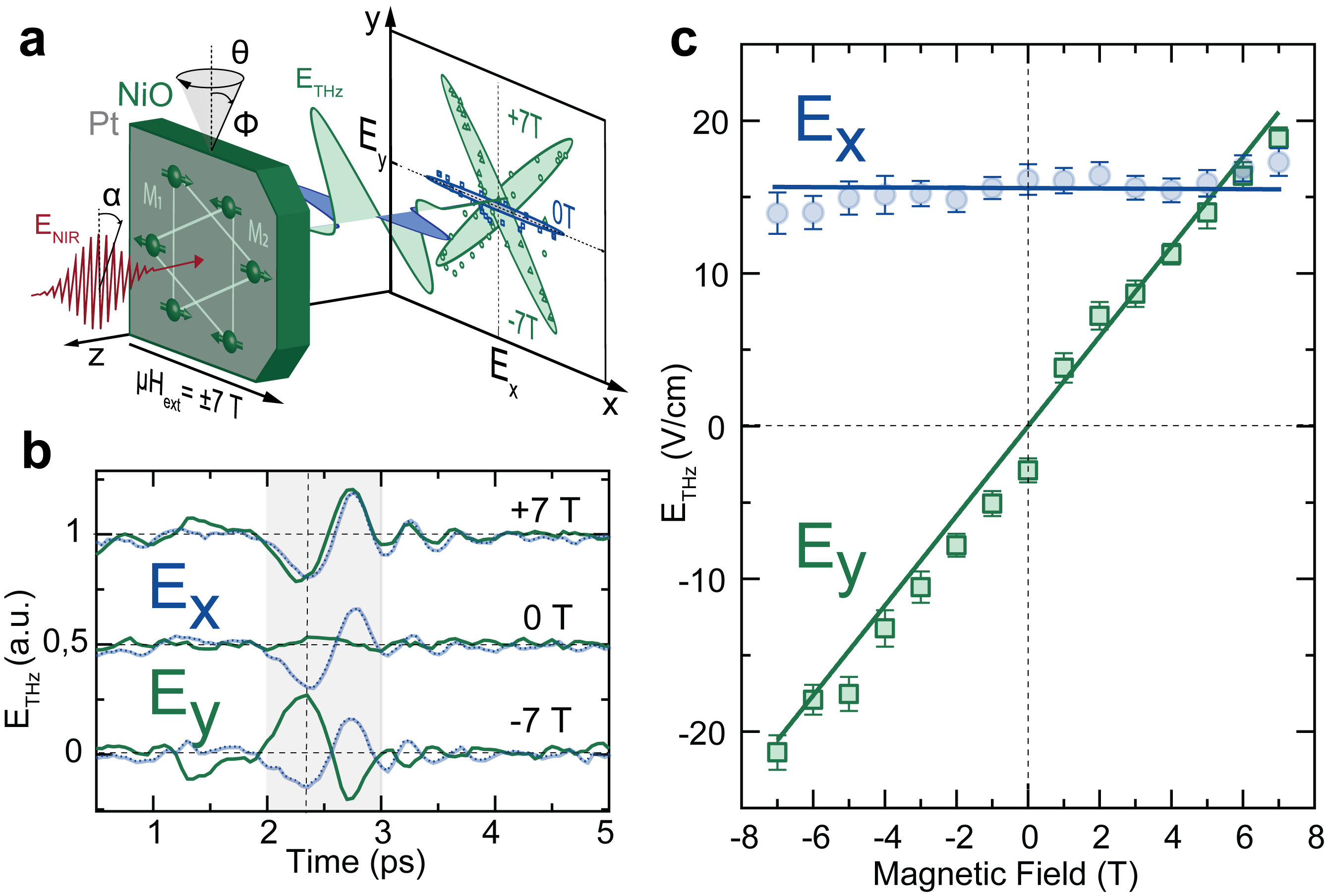}
    \caption{\label{Fig1}
    Separating spin and charge contributions by high magnetic field terahertz polarimetry.
    (a) Experimental geometry for THz polarization sensitive 800~nm pump - THz emission probe spectroscopy of easy-plane (111) NiO heterostructures with laboratory coordinate system. An external magnetic field \Hext is applied along the x-axis, the pump beam is polarized along the y-axis. The rotation of the pump polarization is indicated by an angle $\alpha$ and the sample rotation by the in-plane angle $\phi$ and out-of-plane angle $\theta$. A two-dimensional projection of E\textsubscript{THz} separates a non-magnetic \Ex contribution for \Hext=~0~T and a magnetic \Ey contribution for \Hext$\neq$~0~T. Solid lines serve as guides for the eye.
    (b) Time-domain waveforms for various intensities of the external magnetic field \Hext. The vertical dotted line is situated in the first extreme to highlight the opposite phase contributions of \Ex and \Ey at \Hext=~-7~T. The shaded area is used to extract the peak-peak values. (c) THz electric field peak-peak amplitudes for \Ex and \Ey as function of the external magnetic field. The solid blue line serves as a guide for the eye whereas the green line is a linear fit. Data are shown for $\phi~=~90\degree$ and at T~=~10~K.
    }
\end{figure}

%
Here, we introduce a methodology to separate spin and charge contributions to the THz emission from HM~/~AF heterostructures. In particular, we modify the AF spin ordering by means of an intense external magnetic field. We introduce a rigorous symmetry analysis of the THz polarization that allows to attribute the two distinct components to difference frequency generation and ultrafast laser-induced demagnetization, arising from a canting of the NiO sublattices. 
Our approach consists in investigating Pt~/~NiO(111) heterostructures grown by r.f. magnetron sputtering on a MgO substrate (see end matter B) by means of THz emission spectroscopy. Except otherwise specified, the data presented in the main text are consistently acquired for 10~nm and 5~nm thickness of the NiO and Pt films, respectively. The presence of (111) NiO is confirmed by x-ray diffraction as shown in Fig.~S2(a) of the end matter. Bulk NiO exhibits $m3m$ symmetry above the ordering temperature (T\textsubscript{N}~=~523~K) while it displays a $3m$ translational symmetry and a complex multi-domain structure with a total of 12 twin (T) and spin (S) domains for T~<~T\textsubscript{N}. The magnetic ground state is canonically described in terms of two sublattices $\mathbf{M}_1$ and $\mathbf{M}_2$, aligned antiparallel to each other in the (111) planes~\cite{higuchi_selection_2011,bossini_ultrafast_2021}, see Fig.~\ref{Fig1}(a). 
%

%
For our experiments, we introduce a laboratory coordinate system where the in-plane x-~and y-axes coincide with the crystallographic (111) plane of the NiO film, as shown in Fig.~\ref{Fig1}(a). Laser pulses with a central wavelength of 800~nm and a duration of 35~fs excite the sample, which is situated in a cryostat allowing to vary its temperature in the 1.5~-~300~K range. The sample can be rotated around the z-axis by an angle $\phi$ and over the y-axis by an angle $\theta$. An in-plane DC magnetic field up to \Hext=~7~T is applied. The pump beam is linearly polarized and can be rotated in the xy-plane with respect to the y-axis over an angle $\alpha$. The spot size of the pump beam is set to approximately 
1.5 mm full width at half maximum and the fluence is set to $\approx$~2 mJ/cm\textsuperscript{2}.
The THz light emitted from the sample is analyzed via THz polarimetry to separate the \Ex and \Ey components of the THz electric field and to analyze the three-dimensional distribution in free space. Fig.~\ref{Fig1}(a) shows a two-dimensional projection of the emitted terahertz electric field for \Hext=~-7~T,~0~T,~+~7~T. In the absence of a magnetic field, the electric field of the emitted radiation possesses only an x-component (\EX). A non-vanishing y-component (\Ey) is observed by applying \Hext=~$\pm$~7~T, with comparable amplitude but opposite sign, as the field polarity is reversed. Thus, the ellipticity and rotation of $E_{\textnormal{THz}}$ increase as a function of the intensity of the external magnetic field. The presence of ellipticity implies a phase shift which is indeed observed by the opposite phase of \Ex and \Ey at \Hext=~-~7~T in Fig.~\ref{Fig1}(b). We detect the laser-induced THz emission from the sample by electro-optical sampling in 1~mm thick ZnTe. Further experimental details of the setup are provided elsewhere~\cite{metzger_effect_2022}.
%

%
In Fig.~\ref{Fig1}(b), we show time-domain traces of the THz electric field components \Ex and \Ey for external magnetic fields of \Hext=~-7~T, 0~T, +7~T. The data reveal a distinctly different behaviour of the two THz electric field components as a function of the external magnetic field. This fact is highlighted by the vertical dotted line, which is placed in correspondence with the first extreme of the respective time-domain waveforms. 
Whereas \Ex maintains the same phase, \Ey exhibits a phase reversal upon reversing the sign of the external magnetic field. Moreover, no THz emission is observed for \Ey in the absence of external field. We note that the highly damped oscillations, mostly pronounced in the signal of the \Ex component, have a frequency equal to 2.3~THz (see end matter, Fig.~S1(d)). As this value does not coincide with any eigenfrequency of magnons in NiO, we rule out a magnetic origin of this component of the signal. We note that similar oscillations were reported in literature and ascribed to coherent magnons~\cite{rongione_emission_2023}.
In the following, we focus the discussion on the main waveform of broadband THz emission in the time-domain range between 2 and 3~ps, corresponding to the the gray shaded area in Fig.~\ref{Fig1}(b). We extract the peak-peak value in this time range and plot the \Ex and \Ey components as a function of the external magnetic field in Fig.~\ref{Fig1}(c). The \Ex component is unaffected by the external magnetic field. On the other hand, a linear trend is observed for the \Ey component, which changes sign upon reversal of the field, even exceeding the \Ex component in amplitude. We emphasize that for \Hext=~0~T, no \Ey component is emitted.
%

\begin{figure}
    \centering
    \includegraphics[width=\columnwidth]{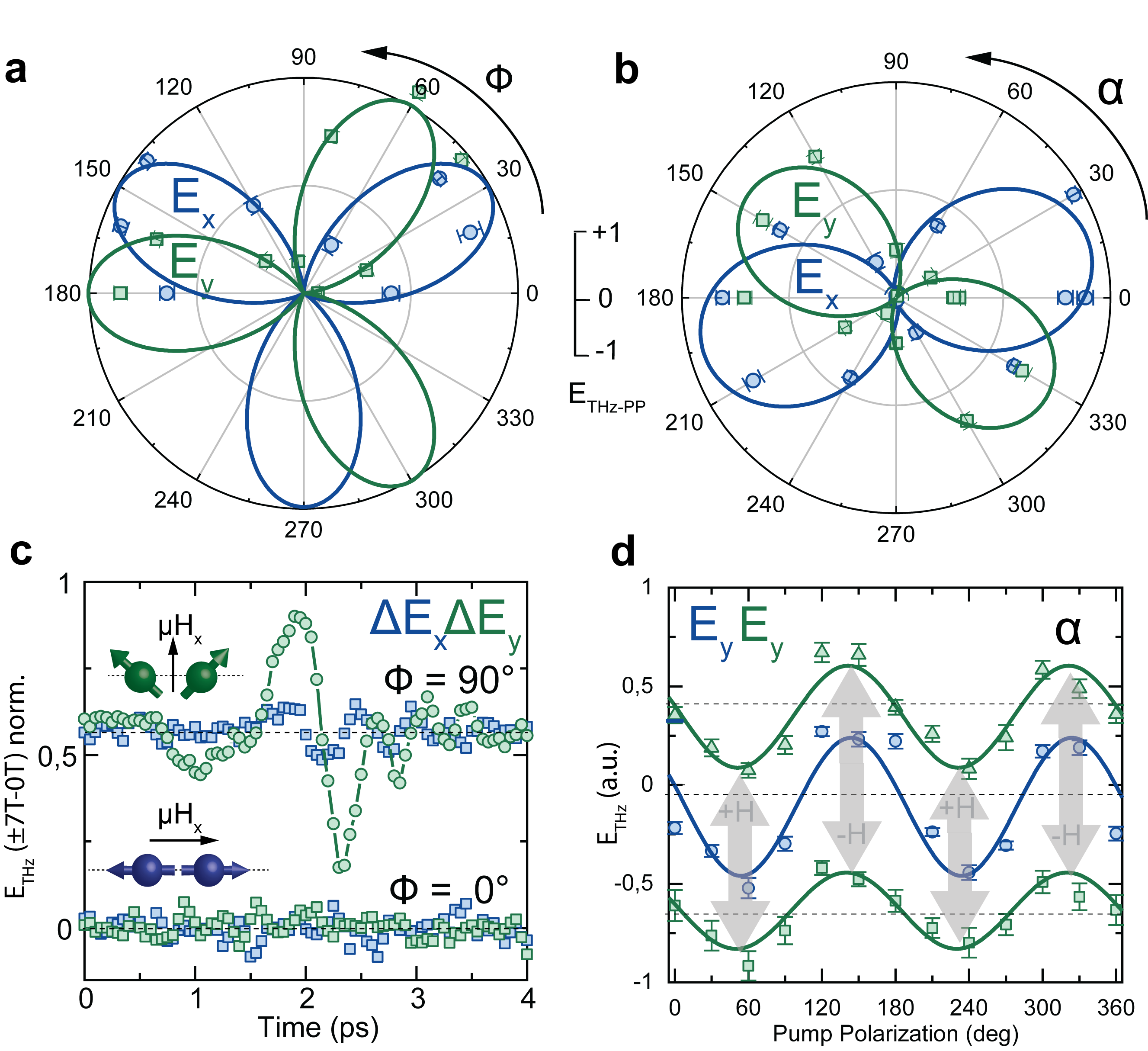}
    \caption{\label{Fig2}
    Separating spin and charge contributions by symmetry. Non-magnetic (a-b) and magnetic (c-d) contributions for an in-plane sample rotation $\phi$ (a,~c) and a pump polarization rotation $\alpha$ (b,~d).
    (a) Sample rotation dependence of \Ex and \Ey at zero field. The solid lines are sinusoidal fits with a periodicity of 120\degree.
    (b) Pump polarization dependence of \Ex and \Ey at zero field. The solid lines are sinusoidal fits with a periodicity of 180\degree.
    (c) Presence (absence) of magnetic field induced THz emission for $\phi$=~90~\degree ($\phi$=~0~\degree) for alignment of \Hext $\bot$ ($\parallel$) spins. The time-domains waveforms of $\Delta$\Ex and $\Delta$\Ey were obtained subtracting data obtained in the absence of external field, from the analogous dataset measured for \Hext=~7~T.
    (d) Presence (absence) of magnetic field induced THz emission for \Hext$\neq$~0~T (\Hext$=$~0~T). The solid lines are sinusoidal fits with a periodicity of 180\degree~ and the in-plane rotation angle is $\phi~=~90\degree$.
    }
\end{figure}

%
We now discuss the origin of the two processes contributing to the THz emission from the Pt~/~NiO heterostructures. We introduce the emitted THz electric field as $\mathbf{E}_{THz}=E_{x}\mathbf{e}_{x}+E_{y}\mathbf{e}_{y}$, where $\mathbf{e}_{x}$ and $\mathbf{e}_{y}$ are the respective unit vectors and \Ex (\Ey) constitutes the magnetic field independent (dependent) THz polarization component as shown Fig.~\ref{Fig1}.
%
%
We start with the magnetic field independent THz component. Several studies~\cite{qiu_ultrafast_2021,rongione_emission_2023} have shown that THz emission can arise from Pt~/~NiO heterostructures without an externally applied magnetic field. One of the mechanisms claimed responsible for THz emission from both bulk and thin film NiO heterostructures is difference frequency generation (DFG)\cite{higuchi_selection_2011,qiu_ultrafast_2021}. Here, two frequency components of electric field of the incident laser pulses \wone and \wtwo generate a third frequency component \wthree=~\wone-~\wtwo via nonlinear light-matter interaction. The DFG process is characterized by a linear fluence dependence, which we indeed observe in our experiment (end matter, Fig.~S1(c)).
The nonlinear optical polarisation responsible for the DFG emission can be expressed as 
$P_{i}(\omega_{1}-\omega_{2})=\chi_{ijk}E_{j}(\omega_{1})E_{k}^{*}(\omega_{2})$. Applying this formalism to our observations, the THz emission is generated by a nonlinear polarization $P_{y}(\omega_{THz})=\chi_{yyx}E_{y}(\omega_{NIR,1})E_{x}^{*}(\omega_{NIR,2})$ with $\chi_{yyx}$, representing the transverse tensor components.
This symmetry condition is verified by our measurements as a function of the in-plane rotation angle $\phi$ for \Hext=~0~T (see Fig.\ref{Fig2}(a). Here, we show that the THz emission amplitude of both \Ex and \Ey changes periodically as a function of $\phi$ with a period of 120\degree. In agreement with earlier observations in bulk~\cite{higuchi_selection_2011} and thin film~\cite{qiu_ultrafast_2021} NiO, we assign this three-fold dependence to the crystallographic \textit{3m} symmetry of NiO below the ordering temperature.
Our experiments as a function of pump laser polarization $\alpha$ reveal a 180\degree~ periodicity of both the \Ex and the \Ey component as shown in Fig.\ref{Fig2}(b). We assign this two-fold dependence to the rhombohedral distortion of the NiO crystal lattice as reported by optical pump probe~\cite{wust_indirect_2022} and optical pump THz emission\cite{qiu_ultrafast_2021} experiments.
We emphasize that our systematic investigation as a function of the external magnetic field, the in-plane sample rotation and the pump polarization allows to conclude that the \Ex component of the THz emission is of purely optical origin. Nevertheless, a comparison of our results with the literature~\cite{qiu_ultrafast_2021,rongione_emission_2023,qiu_terahertz_2023}, demands to address possible contributions of spin transport effects thoroughly. In the first section of the end matter, we address the absence of spin transport effects for \Hext=~0~T. In particular, we stress the absence inverse optical effects (Fig.~S1(b)), the linear fluence dependence (Fig.~S1(c)) and the non-magnetic origin of the highly damped oscillations (Fig.~S1(d)). Ultimately, we rule out a magnetic origin of the \Ex THz polarization component by photoexciting the different sides of the sample. Here, the symmetry of spin transport effects dictates that the THz emission signal must change sign~\cite{qiu_ultrafast_2021,rongione_emission_2023}. However, performing this experiment, we obtain a compelling result: The amplitude of the THz emission is not reversed once different sides of the Pt/NiO sample are optically pumped (Fig.~S1(a)). 
Hence, a magnetic origin of the \Ex component of the laser-driven emission from Pt/NiO heterostructures can be ruled out and we conclude that the origin is of electric dipole nature.
%

\begin{figure}[h!]
     \centering
     \includegraphics[width=\columnwidth]{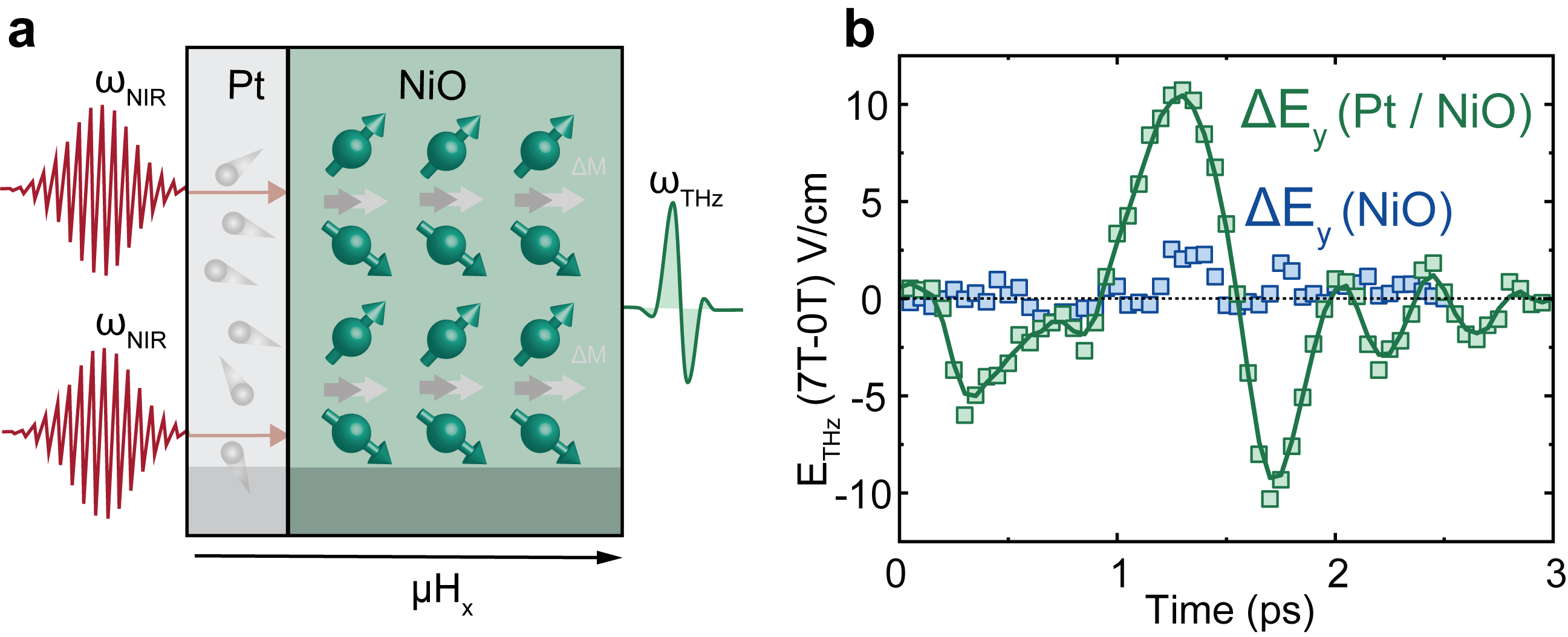}
     \caption{\label{Fig3}\label{Fig4}
     The role of the Platinum for accessing the antiferromagnetic spin system.
     (a) Non-magnetic DFG and magnetic demagnetization processes in HM~/~AFM heterostructures. Non-magnetic DFG occurs independently of the presence of a Platinum layer. Laser-induced demagnetization of the canted NiO spin system occurs via energy transfer of hot electrons in Platinum~\cite{wust_indirect_2022}.
     (b) Comparison of magnetic field induced THz emission from NiO vs Pt~/~NiO heterostructures. The absence of THz emission in the case of NiO emphasizes the crucial role of Platinum as a mediator of magnetic THz emission. The time-domains waveforms were obtained by subtracting the data obtained in the absence of external field from the analogous dataset measured for \Hext=~7~T.
     }
 \end{figure}

%
Differently, the second THz electric field contribution \Ey scales linearly with the fluence (end matter, Fig.~S1(c)), temperature (end matter, Fig.~S2(b)) and the applied magnetic field along the x-axis (Fig.~\ref{Fig1}(c)). These observations are compatible with two possible scenarios. Exploring the framework of electric dipole emission, we can express the nonlinear polarization as $P_{y}=\sum_{i,j}\chi^{e}_{yijx}E_{i}E_{j}H_{x}$, where $i$ and $j$ can be either $x$ or $y$ and $\chi_{yijx}$ are the corresponding elements of the phenomenological fourth-rank tensor. On the other hand, in the case of THz radiation emitted by a magnetic dipole, the magnetization is given by $M_{x}=\sum_{i,j}\chi^{m}_{xijx}E_{i}E_{j}H_{x}$. The former mechanism can be understood as magnetic field-induced difference frequency generation and THz emission from an electric dipole. The latter corresponds to the existence of a field-induced magnetic dipole.
We furthermore observe that the electric dipole contribution implies a term in the energy given by $\mathbf{P} \cdot \mathbf{E}$, which is allowed in a material breaking both time-reversal and space-inversion symmetries. This is not the case for NiO, as it is centrosymmetric. We conclude that the aforementioned electric dipole contribution is symmetry-forbidden in NiO.
Hence, we turn our attention to the magnetic dipole contribution, noting that a magnetic field applied perpendicular to antiferromagnetic spins cants the sublattices $\mathbf{M}_1$ and $\mathbf{M}_2$, effectively generating a magnetic moment $\mathbf{M}~=~\mathbf{M}_{1}~+~\mathbf{M}_{2}~\neq~0$. In this way, a gradual increase of the net magnetization $\Delta$M\textsubscript{x} is observed, as the magnetic field is switched on and its intensity is steadily increased.
A non-zero net magnetic moment M~$\neq$~0 can be a source of THz emission~\cite{seifert_efficient_2016}.
Applying \Hext parallel to the magnetic sublattices, a net magnetization in an AF can only be induced if the field is perfectly aligned with the spins and exceeds a critical value to trigger the spin-flop phase transition~\cite{gurevich_magnetization_1996}. In bulk NiO, this critical value has been estimated to be bigger than 8.5~T~\cite{nogues_isothermal_2003,cottam_effect_1979}. Moreover, a spin flop transition is a first order phase transition corresponding to a rapid increase in the net magnetization contrary to the the aforementioned spin canting~\cite{gurevich_magnetization_1996}.
In Fig.~S3 of the end matter, we show that in the case of \Hext $\bot$ spins, a linear increase of the THz emission signal is detected. Differently, for \Hext $\parallel$ spins, the THz emission is independent of the external magnetic field. Moreover, the continuity of these dependencies confirms that the spin flop field in NiO thin films indeed exceeds \Hext=~7~.
We have already confirmed the three-fold crystallographic symmetry by measuring the emitted THz electric field \Ex and \Ey components, as a function of the sample rotation angle $\phi$ in the absence of an external field (see Fig.\ref{Fig2}(a)). Now, we elaborate on the separation of spin and charge contributions by symmetry and in the presence of a external magnetic field. 
In Fig.\ref{Fig2}(c), we show the magnetic field induced THz emission for in-plane rotation angles of $\phi$~=~0\degree~ and $\phi$~=~90\degree. An externally applied magnetic field has only a negligible effect on the THz emission amplitude of $\Delta$\Ex, which agrees with our conclusion of non-magnetic DFG. On the other hand, $\Delta$\Ey is not detected for $\phi$~=~0~\degree, but surprisingly, it becomes measurable for $\phi$~=~90~\degree.
Thus, we identify this $\Delta$\Ey contribution as picosecond demagnetization of the magnetization following NiO spin canting by external magnetic field $\Delta$\Ey$\propto$~M\textsubscript{x}~$\propto$~\textmu\textsubscript{0}H\textsubscript{x}.
Moreover, the effect of spin canting can also be observed when rotating the pump laser polarization $\alpha$ as shown in Fig.\ref{Fig2}(d). For \Hext=~0~T, a sinusoidal dependence is observed. Applying a field of \Hext=~$\pm$7~T induces a constant offset sensitive to the sign of the externally applied magnetic field.
Assuming that the exchange interaction is kT\textsubscript{N}~=~B\textsubscript{ex}M and the exchange field acting on the spin is B\textsubscript{ex}M, where M is the magnetic moment of the ion, an external magnetic field at the order of the \Neel temperature T\textsubscript{N} in Tesla is required to cant the NiO spins by 90\degree. Thus, we can estimate about 1\degree spin canting in the case of \Hext=$\pm$7~T. This value of spin canting is of the same order as Dzyaloshinskii-Moriya interaction (DMI) induced canting in weak ferromagnets and in agreement with the estimated canting angle for bulk NiO~\cite{machado_spin-flop_2017} at the same field. In Fig.~\ref{Fig1}(c), we quantify the emitted THz electric field strengths of both the non-magnetic (15~V/cm) and the magnetic (20~V/cm) contribution. Remarkably, we observe one order of magnitude larger THz emission amplitudes than in weak ferromagnets~\cite{mikhaylovskiy_terahertz_2015,mikhaylovskiy_selective_2017} and a signal that even exceeds the THz emission amplitude from Co/Pt spintronic emitters with optimized interfaces~\cite{li_thz_2019}. Thus, we can exclude diamagnetic contributions as reported but not quantified in the literature~\cite{qiu_ultrafast_2021}.
In view of all the evidence and discussion presented so far, we ascribe the linear dependence to laser-induced, picosecond demagnetization from canted antiferromagnetic spins. We emphasize that THz polarimetry in combination with external magnetic fields allows to separate the contributions of terahertz electric and magnetic dipole emission from HM~/~AFM heterostructures. This strength of our experimental methodology is essential to identify magnetic contributions and constitutes a crucially needed link compared to earlier THz emission experiments on Pt~/~NiO heterostructures. We illustrate the THz emission processes from Pt~/~NiO heterostructures schematically in Fig.\ref{Fig3}(b) and identify non-magnetic DFG (E\textsubscript{x}) and picosecond demagnetization of canted NiO spins (E\textsubscript{y}).
%

%
Lastly, we discuss the role of platinum for the THz emission processes. The literature recently reported that laser-induced heating of the free electrons in Platinum can induce both coherent~\cite{schmoranzerova_thermally_2023} and incoherent~\cite{wust_indirect_2022} spin dynamics in the magnetic layer of heterostructures. Consistently with these observations, we note that the THz emission measured from pure NiO layers is reduced by a factor two and no magnetic field dependence is observed (see end matter, Fig.~S4). Moreover, no THz emission from pure NiO is observed at room temperature despite previously reported \Neel temperature between 373~K and 533~K in (100)NiO/MgO heterostructures\cite{alders_temperature_1998,baldrati_full_nodate}. In Fig.~\ref{Fig3}(b), we compare the magnetic field induced THz emission signals $\Delta$\Ey(NiO) and $\Delta$\Ey(Pt~/~NiO) at T~=~10~K. Importantly, only in the presence of Platinum, THz emission of magnetic origin is present. This evidence is consistent with the literature reporting the scattering of hot Platinum electrons at the material interface as the source of torque induced on the NiO spins, effectively demagnetizing NiO at picosecond timescales~\cite{wust_indirect_2022}. We note reports of a drastic demagnetization enhancement at thin film surfaces compared to bulk materials~\cite{krieger_ultrafast_2017}. Our data thus demonstrate the crucial role of the heavy metal (Platinum) for accessing the antiferromagnetic (NiO) spin system in ultrathin antiferromagnetic heterostructures in general and in particular to observe magnetic THz emission.
%

%
To conclude, we demonstrate an experimental methodology to separate spin and charge contributions from heavy metal~/~antiferromagnet heterostructures. Taking NiO~/~Pt as representative example, we perform high field terahertz polarimetry measurements to separate magnetic from non-magnetic contributions. We introduce a rigorous symmetry analysis of the terahertz polarization and identify picosecond demagnetization of canted antiferromagnetic spins and difference frequency generation as the magnetic and non-magnetic THz emission mechanisms, respectively. Moreover, we elaborate on the crucial role of the heavy metal as an efficient mediator to access the spin system in antiferromagnetic thin films. Ultimately, our work reveals the highly-debated nature of terahertz light emission from ultrathin antiferromagnetic heterostructures. Our methodology carries great potential as a widely applicable procedure to identify genuine laser-induced magnetic responses from ultrathin heterostructures advancing the established field of spintronics and the emerging concept of orbitronics at THz frequencies.
\begin{acknowledgments}
T.W.J.M. and A.V.K. acknowledge support by the European Union’s Horizon 2020 research and innovation program under the Marie Skłodowska-Curie grant agreement No. 861300 (COMRAD), the European Research Council ERC Grant Agreement No.101054664 (SPARTACUS) and the research program “Materials for the Quantum Age" (QuMat, (registration number 024.005.006) which is a part of the Gravitation program financed by the Dutch Ministry of Education, Culture and Science (OCW).
D.B. acknowledges the support of the DFG programs BO 5074/1-1 and BO 5074/2-1.
T.K. and E.S. are supported by JST CREST (JPMJCR20C1 and JPMJCR20T2), Grant-in-Aid for Scientific Research (Grants No. JP19H05600 and JP24K01326), and Grant-in-Aid for Transformative Research Areas (Grant No. JP22H05114) from JSPS KAKENHI, MEXT Initiative to Establish Next-generation Novel Integrated Circuits Centers (X-NICS) (Grant No. JPJ011438), Japan, and the Institute for AI and Beyond of the University of Tokyo.
\end{acknowledgments}
\bibliography{NiO.bib}
\end{document}